\documentclass[twocolumn,natbib]{svjour3}
\smartqed  
\usepackage{graphicx}
\usepackage{amsmath, amssymb}

\journalname{Biol. Cybern.}
\begin{document}

\title{The firing statistics of Poisson neuron models driven by slow stimuli}


\author{Eugenio Urdapilleta \and In\'es Samengo}

\institute{E. Urdapilleta \at
              Centro At\'omico Bariloche and Instituto Balseiro, San Carlos de Bariloche (8400), R\'io Negro, Argentina\\
              Tel.: +54-2944-445100\\
              Fax:  +54-2944-445299\\
              \email{urdapile@ib.cnea.gov.ar}
           \and
           I. Samengo \at
              Centro At\'omico Bariloche and Instituto Balseiro, San Carlos de Bariloche (8400), R\'io Negro, Argentina\\}

\date{Received: date / Accepted: date}

\maketitle

\begin{abstract}
The coding properties of cells with different types of receptive fields have been studied for decades. ON-type neurons fire in response to positive fluctuations of the time-dependent stimulus, whereas OFF cells are driven by negative stimulus segments. Biphasic cells, in turn, are selective to up/down or down/up stimulus upstrokes. In this paper, we explore the way in which different receptive fields affect the firing statistics of Poisson neuron models, when driven with slow stimuli. We find analytical expressions for the time-dependent peri-stimulus time histogram and the inter-spike interval distribution in terms of the incoming signal. Our results enable us to understand the interplay between the intrinsic and extrinsic factors that regulate the statistics of spike trains. The former depend on biophysical neural properties, whereas the latter hinge on the temporal characteristics of the input signal.


\end{abstract}

\section{Introduction}

\indent Different sensory neurons have different filtering properties. The population distribution of these filtering properties has been most extensively studied in peripheral sensory neurons. For example, in a study carried out by \citep{segev2006}, the receptive fields of ganglion cells in the salamander retina were characterized by their polarity (ON, OFF, biphasic ON or biphasic OFF), and temporal scale (fast, medium, slow). A broad distribution of receptive field types was found. However, not all possible receptive fields were equally represented in this distribution: 93\% of the studied cells clustered into a few well defined prototypes. If there are discrete classes of cells, then presumably the nervous system requires specific pre-processing operations on the external signals. Such operations are carried out by the particular types of receptive fields found in each animal. Here we aim at providing a rigorous characterization of the type of information encoded by single cells, depending on their filtering properties. We work with arbitrary Poisson-like neurons, described by a sequence of Volterra filters. In those cases where the neuronal dynamics reduces to a cascade model, the Volterra filters may be replaced by a linear receptive field, and a static non-linear function governing spike generation. Our aim is to understand how the filtering neuronal properties (captured by the Volterra filters, or alternatively, the receptive field and the non-linearity) determine which stimulus features are represented in the peri-stimulus time histogram (PSTH) of a neuron. In addition, we also study how they affect the inter-spike interval (ISI) distribution. We derive analytical expressions for the PSTH and the ISI distribution, and we discuss how the statistical properties of the responses depend on intrinsic and extrinsic factors, differentiating in particular intrinsic cellular properties from stimulus characteristics.

\indent We work in the limit of slowly varying input signals. Our results are thus only relevant for neurons driven by slow stimuli. Examples can be found in hippocampal interneurons receiving massive theta input from the septum \citep{toth1997} or in olfactory neurons modulated by the respiration cycle \citep{fontanini2006}. An even slower action is exerted by modulatory neurotransmitters in the brain. These signals control the overall level of arousal \citep{saper2000,doya2002}, and they affect the intrinsic neuronal properties of their target neurons in time scales of seconds, that is, much longer than the time scales governing spike dynamics. As a final example, slow adaptation processes can also be described in terms of intrinsic transient currents with long time constants \citep{ermentrout1998,benda2003,wilson2004}, that either increase
(facilitation) or decrease (adaptation) the excitability of the neuron.

\section{Derivation of the PSTH}
\indent In Poisson neurons, the probability of generating a spike at time $t$ conditional to a time dependent stimulus $s$ may be expanded in a Volterra series as

\small
\begin{eqnarray}
\label{eq1}
 & & P[t|s]  =  h_{0} + \int_{-\infty}^{+\infty} {\rm d}t_{1}~h_{1}(t-t_{1})~s(t_{1}) + \\
 & & + \int_{-\infty}^{+\infty} {\rm d}t_{1} \int_{-\infty}^{+\infty} {\rm d}t_{2}~h_{2}(t-t_{1},t-t_{2})~ s(t_{1})~s(t_{2}) + \dots + \nonumber \\
 & & + \int_{-\infty}^{+\infty} {\rm d}t_{1} \dots \int_{-\infty}^{+\infty} {\rm d}t_{k}~ h_{k}(t-t_{1},\dots,t-t_{k})~s(t_{1})\dots s(t_{k}) + \nonumber \\
 & & + \dots \nonumber
\end{eqnarray}

\normalsize
\indent Equation (\ref{eq1}) is valid under the assumption that the firing probability only depends on the stimulus history, and not on the previous activity of the neuron. As a consequence, the kernels $h_{k}$ only depend on the differences $t - t_{i}$, as opposed to separate dependencies on $t$ and $t_{i}$. Thus, within this picture, intrinsic bursting, refractoriness, adaptation, and all response-induced modulation processes are neglected.

\indent The kernels $h_{k}(x_{1},\dots,x_{k})$ quantify the efficacy of $s(-x_{1}), \dots, s(-x_{k})$ in modulating the spiking probability at time $t=0$, and $h_0$ accounts for spontaneous firing. Each $h_{k}$ is measured in units of $[{\rm time}]^{-(k+1)}\cdot [{\rm stimulus}]^{-k}$. Causality imposes that $h_{k}(x_{1},\dots,x_{k}) = 0$, if any $x_{i} < 0$.

\indent Stimuli occurring in the distant past cannot affect the firing probability at present times. Hence, we define $\tau_{\rm m}$ as the memory time of the system, such that all kernels $h_{k}(x_{1},\dots,x_{k}) \approx 0$, if any of the $x_{i}>\tau_{\rm m}$. Hence, the integration limits $(-\infty,+\infty)$ of Eq.~(\ref{eq1}) may be shortened to $(t-\tau_{\rm m},t)$, see Appendix.

\indent A quasi static approximation of the firing statistics is possible when the stimulus $s(t)$ evolves slowly in intervals of size $\tau_{\rm m}$. In this case, inside all the integration symbols we may replace the functions $s(t_{i})$ by their series expansions around the limit of integration $t$. Thus, for $t_i \in [t-\tau_{\rm m}, t]$,

\small
\begin{eqnarray}
\label{eq3}
 s(t_{i}) & = & \sum_{j = 0}^{+\infty} \frac{\lambda_{j}}{j!} ~ (t_{i}-t)^{j}, \ \ \ {\rm where} \\
 \lambda_{j} & = & \left.\frac{{\rm d}^{j} s}{{\rm d}t^{j}}\right|_{t}, \nonumber
\end{eqnarray}

\normalsize 
\noindent and $\lambda_{0} = s(t)$. If the stimulus varies slowly, then Eq.~(\ref{eq3}) is a rapidly converging series inside $[t - \tau_{\rm m}, t]$, so we only keep the first few terms in the expansion. The parameters $\lambda_{j}$ capture the effect of the stimulus on the PSTH. In order to separate this effect from the one produced by intrinsic neuronal properties, we define the coefficients $H_{k}^{j_{1},\dots,j_{k}}$, summarizing the neurons's filtering characteristics, namely,

\small
\begin{eqnarray}
\label{eq6}
H_{k}^{j_{1},\dots,j_{k}} & = & \int_{t-\tau_{\rm m}}^{t} {\rm d}t_{1} \dots \int_{t-\tau_{\rm m}}^{t} {\rm d}t_{k} ~ h_{k}(t-t_{1},\dots,t-t_{k}) \prod_{i = 1}^{k} (t_{i}-t)^{j_{i}} \nonumber \\
 & = & \int_{0}^{\tau_{\rm m}} {\rm d}x_{1} \dots \int_{0}^{\tau_{\rm m}} {\rm d}x_{k} ~ h_{k}(x_{1},\dots,x_{k}) \prod_{i = 1}^{k} (-x_{i})^{j_{i}}.
\end{eqnarray}

\normalsize
\indent Replacing Eq.~(\ref{eq3}) in (\ref{eq1}) and using the definition (\ref{eq6}), the probability $P[t|s]$ is expanded as a series of terms that only depend on time through the instantaneous value of the stimulus and its derivatives at time $t$

\small
\begin{equation}
\label{eq7}
P[t|s] = P_{0}[s(t)] + P_{1}[s(t),s'(t)] + P_{2}[s(t),s'(t),s''(t)] + \dots ,
\end{equation}

\normalsize
\noindent where, explicitly,

\small
\begin{equation}
\label{eq8}
P_{r}[s,s',\dots,s^{(r)}] = h_{0} ~ \delta_{r,0} + \sum_{k = 1}^{+\infty} \sum'_{\begin{array}{c}\{j_{1},\dots, j_{k}\} / \\ \sum_{i = 1}^{k} j_{i} = r \end{array}} \quad \prod_{\ell = 1}^{k} \frac{\lambda_{j_{\ell}}}{j_{\ell}!} ~ H_{k}^{j_{1},\dots,j_{k}},
\end{equation}

\normalsize
\noindent and the primed sum over the $\{j_{1},\dots,j_{k}\}$ runs through all non-negative integers that sum up to $r$, that is, $\sum_{i = 1}^{k} j_{i} = r$ (see Appendix for a detailed derivation).

\indent The value of $H_{k}^{j_{1},\dots,j_{k}}$ is only dependent on intrinsic neuronal properties (the filters $h_{k}$), whereas the parameters $\lambda_{i}$ incorporate the effect of the stimulus $s(t)$.

\indent For example, the first two terms read 

\small
\begin{eqnarray}
\label{eq9}
P_{0}(s) & = & h_{0} + \sum_{k = 1}^{+\infty} \left(\lambda_{0}\right)^{k} ~ H_{k}^{00 \dots 0}, \\
\label{eq10}
P_{1}(s) & = & \sum_{k = 1}^{+\infty} \left(\lambda_{0}\right)^{k - 1} \lambda_{1} \left[H_{k}^{10 \dots 0} + H_{k}^{01 \dots 0} + \dots + H_{k}^{00 \dots 1} \right].
\end{eqnarray}

\normalsize 
\indent Notice that $P_{0}(s)$ is the activation curve: the function that relates the firing rate of the cell to the strength of a constant stimulus $s$. In Eq.~(\ref{eq9}), this activation curve is written as a Taylor expansion in $s$. For arbitrary $r$, each $P_{r}$ depends on the instantaneous value of the stimulus and its derivatives: $P_{0}$ depends only on $s(t)$, $P_{1}$ depends linearly on $s'(t)$, $P_{2}$ combines quadratic terms in $s'(t)$ with linear terms on $s''(t)$, and so forth. Therefore, if the stimulus varies slowly in intervals of duration $\tau_{\rm m}$, the spiking probability $P[t|s]$ of Eq.~(\ref{eq7}) becomes an instantaneous function of time. This contrasts with the original functional of Eq.~(\ref{eq1}), containing the whole stimulus history.

\subsection{Application to linear neuron models}

\indent In a linear neuron model, the firing probability reads

\small
\begin{equation}
\label{eq11}
P[t|s] = h_{0} + \int_{t-\tau_{\rm m}}^{t} h_{1}(t-t_{1})~s(t_{1})~dt_{1}.
\end{equation}

\normalsize
\indent In this case, the only non-vanishing filters are $h_{0}$ and $h_{1}$, so $H_{k}^{j_{1},\dots,j_{k}}$ is zero for all $k > 1$. Hence, the expansion~(\ref{eq7}) reads

\small
\begin{equation}
\label{eq12}
\begin{array}{l}
P_{0}[t|s] = h_{0} + \lambda_{0} ~ H_{1}^{0} \\
P_{1}[t|s] = \lambda_{1} ~ H_{1}^{1} \\
P_{2}[t|s] = \lambda_{2} ~ H_{1}^{2} / 2 \\
\vdots \\
P_{k}[t|s] = \lambda_{k} ~ H_{1}^{k} / k!
\end{array}
\end{equation}

\begin{figure}[h]
\begin{center}
\includegraphics[scale = 0.35, angle = 0]{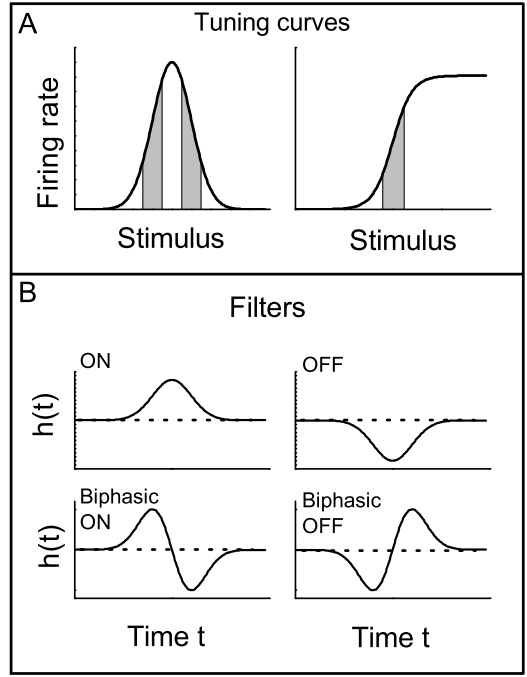} \caption{\label{f1} {\bf Linear neuron models.} {\em A}: Typical tuning curves. The shaded areas represent the range of stimuli where a linear approximation is valid. {\em B}: Typical linear filters $h_1(t)$ representing ON, OFF, and biphasic (ON and OFF) cells. Spike generation results from a convolution of $h_{1}(t)$ with the stimulus. The dotted line represents $h_{1} = 0$}
\end{center}
\end{figure}

\normalsize
\indent A linear model is only applicable within a fairly narrow range of stimulus fluctuations, given that the firing probability must always remain linear and non-negative. In Fig.~\ref{f1}A we show two typical activation curves, or tuning curves. On the left we see a bell-shaped tuning curve, characteristic, for example, of direction and orientation-selective cells in the middle temporal visual cortex \citep{albright1984}, or of motoneurons, selective to the direction of arm-reaching movements \citep{georgopoulos1982}. On the right, the sigmoid-shaped function exemplifies the behavior of many rectifying input-output relations, found for instance in the constrast selective cells in the striate cortex of mammals \citep{albrecht1982}. These curves are obviously non-linear. In both cases, for small and large stimuli, the firing rate saturates, making subtle stimulus discrimination impossible. Hence, a linear approximation of the intrinsic neuronal dynamics is only valid for stimuli that fluctuate around the shaded regions, where the activation curve is well described by a straight line. Although linear models cannot be used outside this restricted range, in many applications they still prove to be useful to capture the temporal properties of the cell's receptive field. In Fig.~\ref{f1}B we show four typical examples of filters $h_{1}(t)$ corresponding to ON, OFF, biphasic ON and biphasic OFF cells, as reported for example by \citep{segev2006}, or \citep{gollisch2008a,gollisch2008b}.

\begin{figure*}
 \centering
   \begin{minipage}[t]{.25\textwidth}
     \centering
     \par\vspace{0pt}
     \caption{\label{f2} {\bf Filters and PSTH of three linear ON and three linear OFF model neurons}. In the PSTH, gray smooth lines correspond to Eq.~(\ref{eq11}), and black irregular traces are from multitrial Poisson simulations. In cells 2 and 5, the amplitude of the filters is smaller than in cells 1 and 4. In cells 3 and 6, the center of mass of the filters is displaced to the right. In ON cells, the PSTH is proportional to the delayed stimulus, whereas in OFF cells it is proportional to the delayed negative stimulus. The proportionality constant depends on the total area below the filter. Thus, in cells 2 and 5, the PSTH has a smaller standard deviation than in cells 1 and 4 (compare the vertical bars to the right of the PSTH). The delay with respect to the stimulus is equal to the location of the center of mass of the filters. Thus, in cells 3 and 6 the delay is larger than in cells 1 and 4 (compare the slope of the lines connecting extremes in the stimulus and the PSTH. In all cases, $h_{0} = 100~{\rm Hz}$ and the PSTH is measured in [${\rm ms}^{-1}$]}
   \end{minipage}%
   \begin{minipage}[t]{.75\textwidth}
     \centering
     \par\vspace{0pt}
     \includegraphics[width=0.75\textwidth]{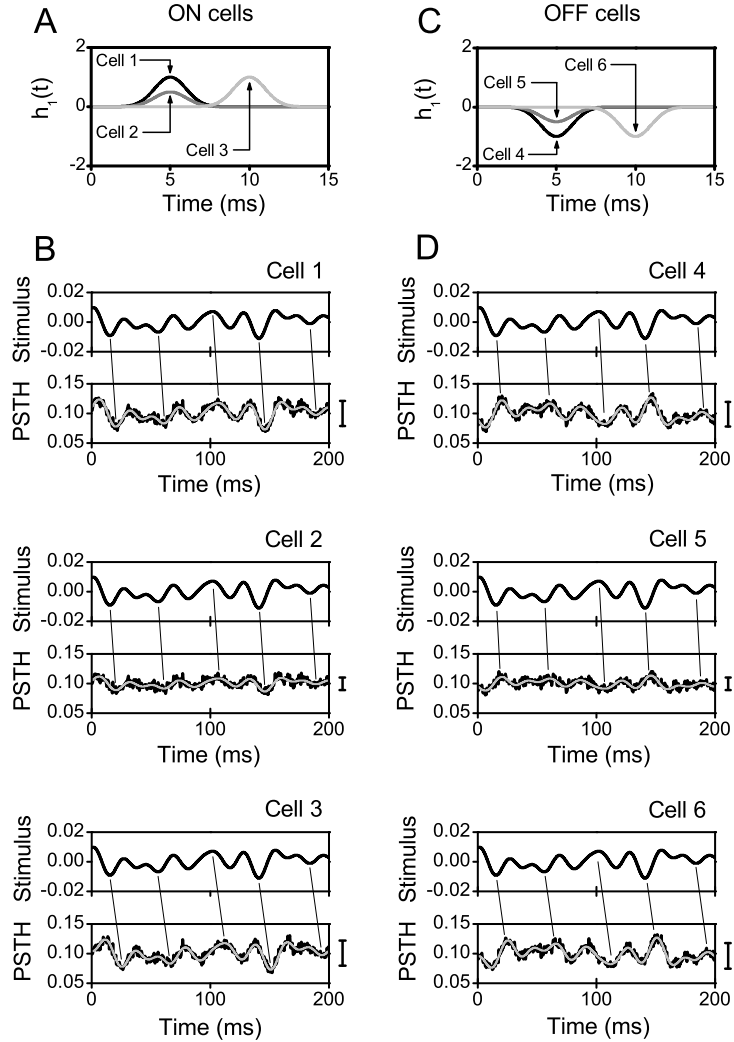}
   \end{minipage}
\end{figure*}

\indent In order to remain in the linear portion of the activation curve, the parameters $h_{0}$ and $H_{1}^{k}$ must fulfill

\small
\begin{equation}
\label{eq13}
   h_{0} \gg \sum_{k = 0}^{+\infty} |H_{1}^{k}| \frac{\sigma_{\lambda_{k}}}{k!}
\end{equation}

\normalsize
\noindent where $\sigma_{\lambda_{k}}$ is the standard deviation of the $k$-th derivative of the time-dependent stimulus.

\indent From Eq.~(\ref{eq12}) we see that $P_{0}[s(t)]$ follows the stimulus linearly, whereas $P_{1}[s(t), s'(t)]$ introduces a correction that is proportional to the derivative of the stimulus. For the time being, we assume that the higher-order terms are negligible. Whenever $|H_{1}^{0}| \gg |H_{1}^{1}| / \tau_{\rm m}$,

\small
\begin{eqnarray}
\label{eq14}
P[t|s] & \approx & h_{0} + H_{1}^{0} s(t) + H_{1}^{1} s'(t) \nonumber \\
       & & h_{0} + H_{1}^{0} \left[ s(t) + \frac{H_{1}^{1}}{H_{1}^{0}} s'(t) \right] \nonumber \\
       & \approx & h_{0} + H_{1}^{0} s(t - \delta_{0}),
\end{eqnarray}

\normalsize
\noindent where $\delta_{0} = - H_{1}^{1} / H_{1}^{0}$. The equality between the last two lines in Eq.~(\ref{eq14}) is
obtained from the first two orders of a Taylor expansion of $s(t -\delta_{0})$. Therefore, if the linear filter $h_{1}(t)$ is such that $|H_{1}^{1}| / |H_{1}^{0}| \tau_{\rm m} \ll 1$, then the PSTH follows the stimulus linearly, with a delay

\small
\begin{equation}
\label{eq15}
\delta_{0} = - \frac{H_{1}^{1}}{H_{1}^{0}} = - \frac{\int_{0}^{\tau_{\rm m}} h_{1}(t) ~ (-t) ~ {\rm d}t}{\int_{0}^{\tau_{\rm m}} h_{1}(t) ~ {\rm d}t} = \frac{\int_{0}^{\tau_{\rm m}} h_{1}(t) ~ t ~ {\rm d}t}{\int_{0}^{\tau_{\rm m}} h_{1}(t) ~ {\rm d}t}.
\end{equation}

\normalsize
\indent The delay $\delta_{0}$ is equal to the center of mass of the filter $h_{1}(t)$. Causality imposes that $\delta_{0}$
be always positive.

\begin{figure*}[ht]
\begin{center}
\includegraphics[width=0.65\textwidth]{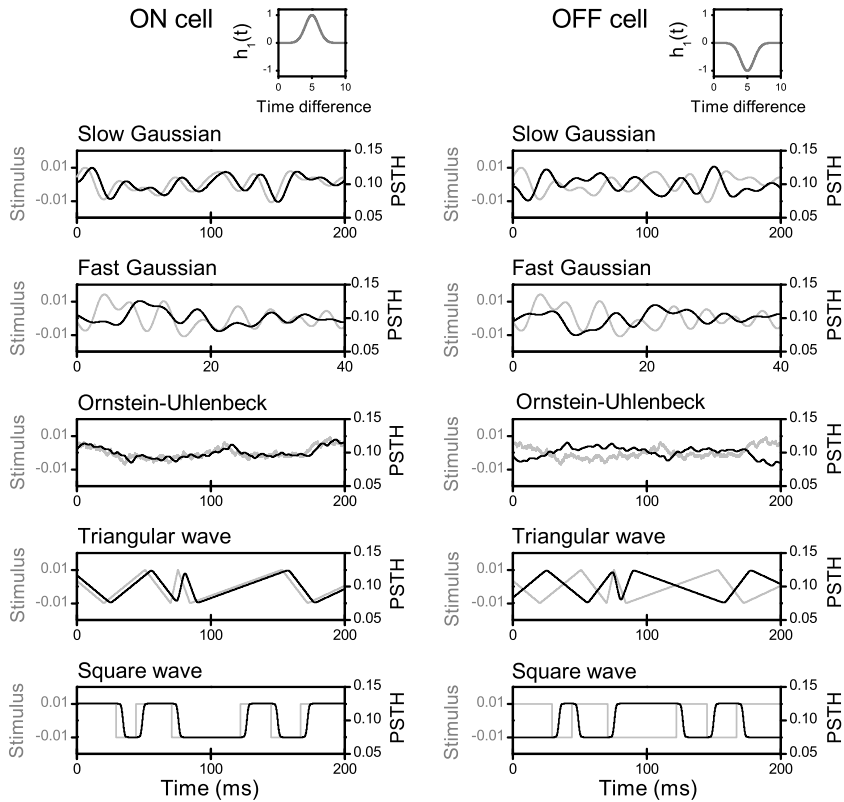}
\caption{\label{f3} {\bf Comparison between several stimuli and the associated PSTH, for an ON and an OFF linear neuron.} Left/right scales measure the stimulus (gray)/PSTH (black). Slow Gaussian stimulus: cutoff frequency 50 Hz. Fast Gaussian stimulus: cutoff frequency 250 Hz. Ornstein-Uhlenbeck process: correlation time 100 ms. Triangular and square stimuli: up- and downstrokes are determined by a Poisson process of mean 25 Hz. Both ON and OFF cells follow the stimulus accurately, except at discontinuous points, where an integration roundoff of approximately $5$ ms is observed. Notice that the $x$-scales vary from panel to panel. In all cases, $h_{0} = 100~{\rm Hz}$ and the PSTH is measured in [${\rm ms}^{-1}$]}
\end{center}
\end{figure*}

\indent If the filter $h_{1}$ is always positive (i.e., the cell responds to depolarizing stimuli, as the ON filter in
Fig.~\ref{f1}B), then $H_{1}^{0}$ is positive, and $H_{1}^{1}$ is negative. The firing probability is then proportional to the delayed stimulus, as exemplified in cells 1-3, in Fig.~\ref{f2} A and B. The constant of proportionality between the PSTH and the stimulus is equal to $H_{1}^{0}$, that is, the area below the filter $h_{1}(t)$. As the amplitude of $h_{1}(t)$ decreases (compare cell 2 with cell 1, in Fig.~\ref{f2}), the standard deviation of the resulting PSTH also decreases (compare the vertical bars to the right of the PSTHs). In turn, the delay $\delta_{0}$ is equal to the center of
mass of $h_{1}(t)$. Therefore, if $h_{1}(t)$ is displaced to the right (compare cell 3 with cell 1), then the time-lag between the stimulus and the PSTH increases (compare the slope of the lines connecting corresponding points in the stimulus and the PSTH).

\indent If $h_{1}$ is always negative (i.e., the cell responds to hyperpolarizing stimuli, as the OFF cells 4-6 in Fig.~\ref{f2} C and D), then $H_{1}^{0}$ is negative and $H_{1}^{1}$ is positive. Thus, the firing probability is proportional to the delayed negative stimulus, as shown in Fig.~\ref{f2}. As the amplitude of the filter diminishes (compare cell 5 with cell 4), the standard deviation in the PSTH decreases accordingly. If the filter is delayed (see cell 6), the time-lag between the stimulus and the PSTH increases as well.

\indent It is important to notice that the amplitude of the PSTH is {\sl only} determined by $H_{1}^{0}$, and the delay $\delta_{0}$ {\sl only} depends on the center of mass of the filter. Other modifications of $h_{1}(t)$ that preserve its total area will not change the proportionality constant between the stimulus and the PSTH. In a similar way, distortions that do not modify the center of mass of the filter, leave the time-lag unchanged.

\indent In Fig.~\ref{f3}, the PSTH of an ON and an OFF linear neuron is compared to the injected stimulus. Five different types of stimuli are used. For both cells the delay $\delta_{0}$ is equal to 5 ms. For slow enough stimuli the PSTH is equal to the delayed stimulus (or negative stimulus) with a time-lag of 5 ms (see slow Gaussian in Fig.~\ref{f3}). As the stimulus varies faster (see fast Gaussian in Fig.~\ref{f3}), the PSTH becomes a smoothed version of the delayed stimulus. In the case of the Ornstein-Uhlenbeck process, only high frequencies are filtered out. Thus, for the triangular wave, the sharp edges of the stimulus at each switching point are converted into 5 ms rounded corners in the PSTH (5 ms is also the total width of the kernel $h_{1}$). Similarly, the discontinuous square wave produces a continuous PSTH, where the rise and the decay times between the up and down states also last for approximately 5 ms.

\begin{figure*}[ht]
\begin{center}
\includegraphics[width=0.65\textwidth]{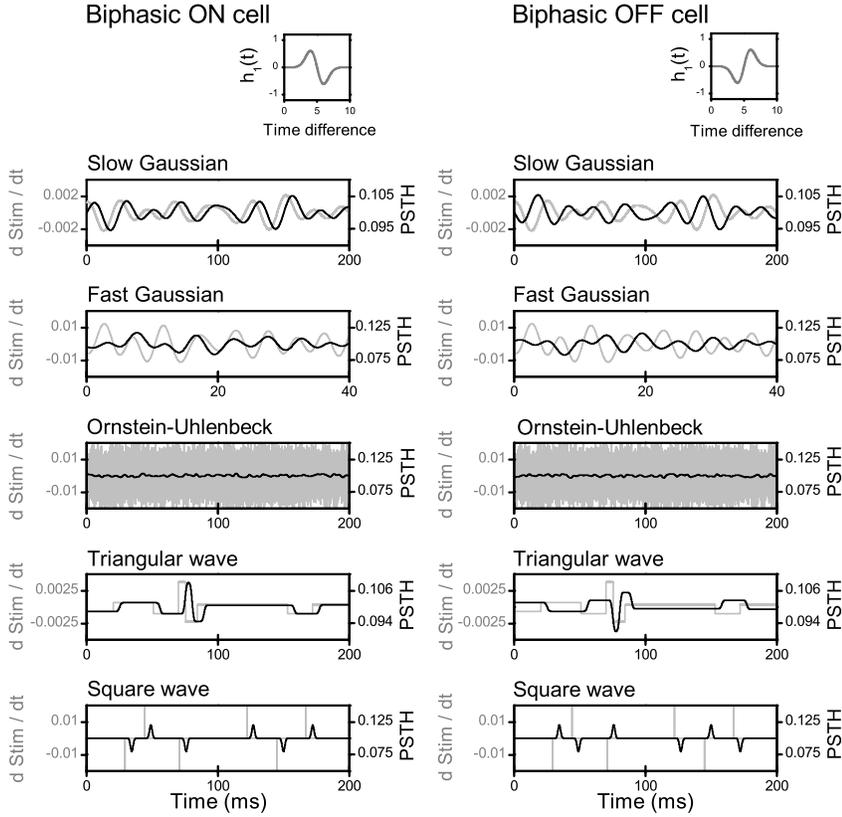} \caption{\label{f4}
{\bf Comparison between several stimuli and the associated PSTH, for an ON-biphasic and an OFF-biphasic linear neuron.} Left/right scales measure the stimulus derivative (gray)/PSTH (black). Slow Gaussian stimulus: cutoff frequency 50 Hz. Fast Gaussian stimulus: cutoff frequency 250 Hz. Ornstein-Uhlenbeck process: correlation time 100 ms. Triangular and square stimuli: up- and downstrokes are determined by a Poisson process of mean 25 Hz. Both the biphasic-ON and the biphasic-OFF cells follow the derivative of the stimulus accurately, except at discontinuous points. The Ornstein-Uhlenbeck process has a discontinuous derivative, so the quasi-static approach is not applicable. The derivative of a triangular wave is a collection of constant segments. The derivative of a square wave is a sequence of delta-like pulses. Notice that the $x$-scale varies from panel to panel. In all cases, $h_{0} = 100~{\rm Hz}$ and the PSTH is measured in [${\rm ms}^{-1}$]}
\end{center}
\end{figure*}

\indent When $h_{1}$ has both depolarizing and hyperpolarizing phases (see, for example, the biphasic filters in Fig.~\ref{f1}B), then the condition $|H_{1}^{0}| \gg |H_{1}^{1}| / \tau_{\rm m}$ is less likely to hold. In the extreme case of a perfectly symmetric biphasic filter $H_{1}^{0}$ vanishes. The fluctuations in the PSTH are no longer proportional to the delayed stimulus. In fact, whenever $|H_{1}^{1}| \gg |H_{1}^{2}| / \tau_{\rm m}$,

\small
\begin{eqnarray}
\label{eq16}
P[t|s] & \approx & h_{0} + H_{1}^{1} ~ s'(t) + H_{1}^{2} ~ s''(t) \nonumber \\
       & & h_{0} + H_{1}^{1} \left[ s'(t) + \frac{H_{1}^{2}}{H_{1}^{1}} ~ s''(t) \right] \nonumber \\
       & \approx & h_{0} + H_{1}^{1} s'(t - \delta_{1}),
\end{eqnarray}

\normalsize
\noindent where

\small
\begin{equation}
\label{eq17}
\delta_{1} = - \frac{H_{1}^{2}}{H_{1}^{1}} = \frac{\int_{0}^{\tau_{\rm m}} ~ h_{1}(t) ~ t^{2} ~ {\rm d}t} {\int_{0}^{\tau_{\rm m}} ~ h_{1}(t) ~  t ~ {\rm d}t}.
\end{equation}

\normalsize
\indent Thus, in this case, the PSTH follows the first derivative of the stimulus, with a proportionality constant equal to
$H_{1}^{1}$ and a time lag given by $\delta_{1}$, as exemplified in Fig.~\ref{f4}. For the slow Gaussian stimulus, the PSTH is proportional to the derivative of the delayed stimulus. When the stimulus varies faster, as in the fast Gaussian stimulus, the PSTH smoothes rapid fluctuations out. In the limit case of the Ornstein-Uhlenbeck process, the derivative of the stimulus is discontinuous, and the quasi-static approximation is not applicable. Therefore, the black curve in Fig.~\ref{f4} deviates significantly from the derivative of the stimulus (in grey: white noise realization). For triangular and square stimuli, the PSTH detects jumps in the derivative. In the case of square waves, the derivative of the stimulus is a sequence of delta-functions. Although the quasi-static approximation is not applicable in this case either, Fig.~\ref{f4} shows that the model captures the transition times correctly.

\subsection{Application to linear-nonlinear neuron models}

\indent Consider now a neuron model where the probability to generate a spike at time $t$ is

\small
\begin{equation}
\label{eq18}
P[t|s] =  g\left[ h_{0} + \int_{t-\tau_{\rm m}}^{t} h_{1}(t-t_{1})~s(t_{1})~{\rm d}t_{1} \right],
\end{equation}

\normalsize

\begin{figure*}[ht]
\begin{center}
\includegraphics[width = 0.75\textwidth]{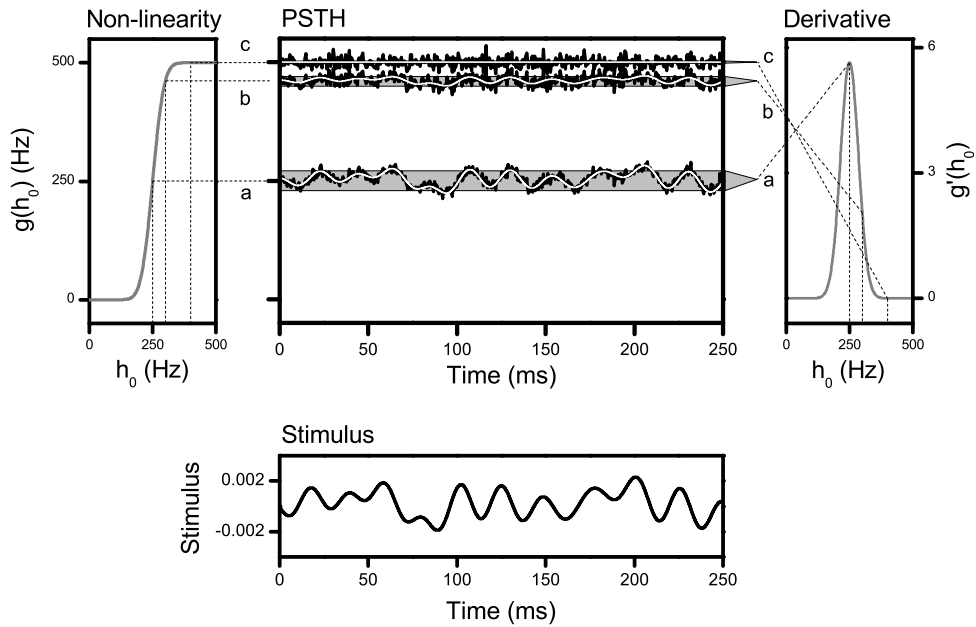} \caption{\label{f5} {\bf The PSTH of linear-nonlinear neuron models}. A slow Gaussian stimulus of $50$ Hz cutoff frequency and $10^{-3}$ standard deviation is presented to a linear-nonlinear model with the same ON filter as cell 1 in Fig.~\ref{f2} ($H_{1}^{0}=2.506$). Left: Nonlinear function $g(h_{0})$. Right: $g'(h_{0})$. Middle: PSTH obtained from numerical simulations (black line) and from Eq.~\ref{eq18} (white line). The three cases {\sl a}, {\sl b}, and {\sl c} correspond to $h_{0} = 250$, $300$ and $400$ Hz, respectively. The mean PSTH is determined by $g(h_{0})$, as indicated by the dotted lines connecting the left and middle panels. Its SD (represented by the height of the grey areas in the middle panel) depends on $g'(h_{0})$, as indicated by the dotted lines connecting the middle and the right panels. The delay with respect to the stimulus is independent of $h_{0}$ and $g$}
\end{center}
\end{figure*}

\noindent where $g$ is a static, nonlinear function. These models have been widely used in vision \citep{rust2005,schwartz2006,brenner2000,gollisch2008a,gollisch2008b} and audition \citep{nagel2006,lesica2008}. Equation~(\ref{eq18}) is not expressed as a Volterra series, because the static nature of $g$ allows us to work with a compact expression, with no need to expand in powers of the stimulus. With this model, when considering an ON (OFF) cell, if Eq.~(\ref{eq13}) holds,

\small
\begin{eqnarray}
\label{eq19}
P[t|s] & = & g\left[h_{0} + \sum_{k=0}^{+\infty} H_{1}^{k} \frac{\lambda_{k}}{k!}\right] \nonumber \\
       & \approx & g\left[h_{0} + H_{1}^{0} ~ s(t-\delta_{0}) \right] \nonumber  \\
       & \approx & g(h_{0}) + g'(h_{0}) ~ H_{1}^{0} ~ s(t-\delta_{0}) + \nonumber \\
       & & \hspace{2.0cm} + \frac{1}{2} ~ g''(h_{0}) ~ \left[ H_{1}^{0} ~ s(t-\delta_{0}) \right]^{2} .
\end{eqnarray}

\normalsize
\indent If we only consider up to the linear term, the similarity between Eqs.~(\ref{eq19}) and (\ref{eq12}) implies that
linear-non linear neuron models are formally analogous to purely linear models: in ON or OFF cells, the PSTH follows the stimulus linearly with a fixed time lag. An analogous argument can be developed for symmetric biphasic neurons, showing that the PSTH follows the stimulus' derivative. The only difference with the purely linear case relies on the constant terms $g(h_{0})$ and $g'(h_{0})$: now the proportionality constant between the PSTH and the stimulus (or its derivative) depends on the operation level $h_{0}$. The non-linear function $g$, nevertheless, has no effect on the time lags $\delta_{0}$ and $\delta_{1}$, since the latter depend only of the ratios $H_{1}^{1}/H_{1}^{0}$ and $H_{1}^{2}/H_{1}^{1}$.

\indent In Fig.~\ref{f5} we see the PSTH of a linear-nonlinear neuron model (central panel), for three different operation levels $h_{0}$. The three $h_{0}$ values can either correspond to three different cells with different spontaneous rates, or to one same cell driven with three stimuli with different mean values. As $h_{0}$ is near saturation, the average PSTH approaches its maximum possible value. In addition, the derivative $g'(h_{0}) \to 0$, therefore the variance of the PSTH decreases.

\begin{figure*}
 \centering
   \begin{minipage}[t]{.25\textwidth}
     \centering
     \par\vspace{0pt}
     \caption{\label{f6}
     {\bf The effect of the concavity of the activation curve.} Upward or downward stimulus fluctuations may or may not be intensified in the PSTH, depending on the concavity of the non-linear function $g$, at the operation level $h_{0}$. The symbols $\oplus$ and $\ominus$ indicate the contribution of the linear and the quadratic terms in Eq.~(\ref{eq19}). When both contributions have the same sign, the feature is enhanced}
   \end{minipage}%
   \begin{minipage}[t]{.75\textwidth}
     \centering
     \par\vspace{0pt}
     \includegraphics[width=0.75\textwidth]{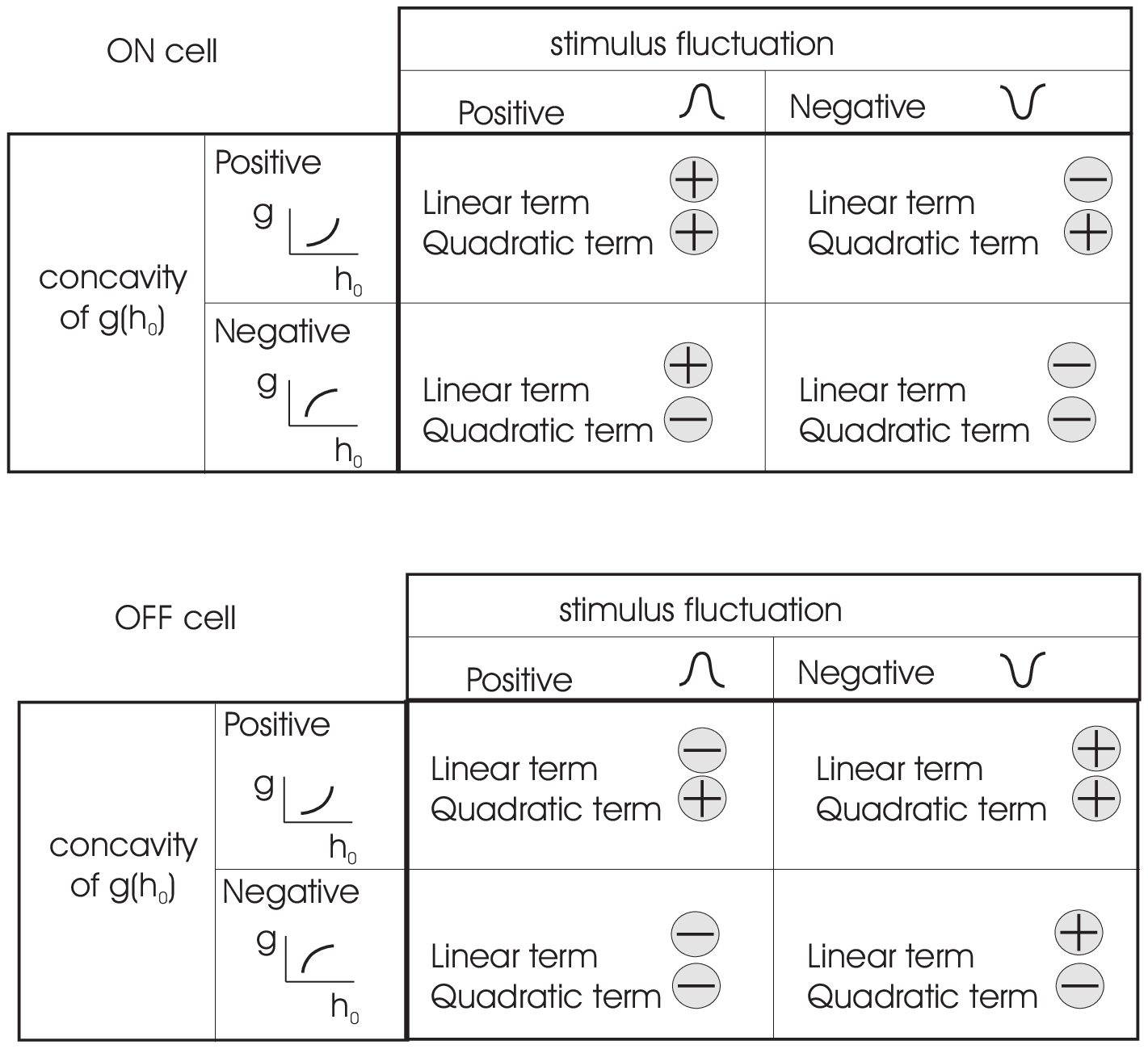}
   \end{minipage}
\end{figure*}

\begin{figure*}[ht]
\begin{center}
\includegraphics[width=0.75\textwidth]{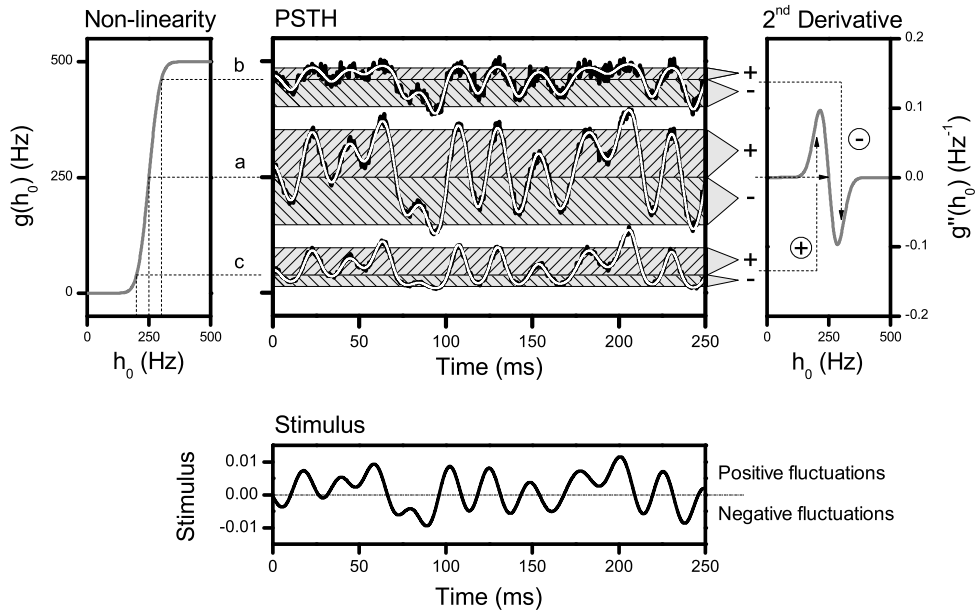} \caption{\label{f7} {\bf Deviations in the PSTH of linear-nonlinear neuron models, for stimuli with large SD}. A slow Gaussian stimulus of $50$ Hz cutoff frequency is presented (see bottom panel). The SD of the stimulus is $5 {\rm x} 10^{-3}$. The linear-nonlinear neuron model has the same ON filter as cell 1 in Fig.~\ref{f2} ($H_{1}^{0}=2.506$). Left panel: Nonlinear function $g(h_{0})$. Three different operation points $h_{0}$ are selected. The corresponding values of $g(h_{0})$ set the spontaneous firing rate, as shown by the thin connecting lines joining the left and middle panels. In case (a), $g(h_{0})$ is at its inflection point, so the concavity $g''(h_{0}) = 0$. In (b), $g''(h_{0}) < 0$, whereas in (c), $g''(h_{0}) > 0$. The resulting PSTHs are displayed in the middle panel, line conventions as in Fig.~\ref{f5}. In case (a), the upward and downward fluctuations in the stimulus are symmetrically represented in the PSTH. Instead, positive fluctuations are flattened in (b), whereas negative fluctuations are amplified. The opposite effect is observed in (c). The differential amplification of positive and negative fluctuations can be predicted from the value of the second derivative of $g$, as shown by the thin connecting lines to the right panel}
\end{center}
\end{figure*}

\indent We now discuss the third term in Eq.~(\ref{eq19}). This contribution is not present in the purely linear neuron model, and is proportional to the squared stimulus, that is, the energy of the input signal. Its effect becomes significant for fairly large stimulus amplitudes. The quadratic term may or may not have the same sign as the linear term, depending on several factors. The scheme in Fig.~\ref{f6} summarizes the combined effects of stimulus fluctuations (upward or downward) and the concavity of $g$ at the operation level $h_{0}$, both for the linear and quadratic terms of Eq.~(\ref{eq19}). For example, if $g''(h_{0}) > 0$, an ON cell transforms positive stimulus fluctuations in enhanced peaks in the PSTH. In this case, both the linear and the quadratic terms contribute with the same sign. Instead, negative fluctuations appear less pronounced, since the linear and quadratic terms have opposite signs. As a consequence, the PSTH is no longer symmetric with
respect to its mean value: upward fluctuations are amplified, and downward fluctuations appear as less pronounced. Instead, if $h_{0}$ is such that $g''(h_{0}) < 0$, positive stimulus fluctuations are flattened, whereas negative ones appear as enhanced troughs. These effects deviate from the purely linear prediction, as exemplified in Fig.~\ref{f7}. In case (a), the operation point $h_{0}$ is exactly at the inflection point of $g(h_{0})$. Therefore, the third term in Eq.~(\ref{eq19}) vanishes, and the fluctuations in the stimulus are represented symmetrically in the PSTH. However, if $h_{0}$ is such that $g(h_{0})$ has negative concavity (case (b)), negative fluctuations are amplified, whereas positive ones are flattened (see
Fig.~\ref{f6}). This asymmetry is clearly visible from the different heights of the two striped grey areas representing the amplitude of positive and negative fluctuations, in the top PSTH of the middle panel. The opposite effect (amplified upward fluctuations and flattened downward fluctuations) is observed in case (c).

\section{The inter-spike interval distribution}

\indent In Poisson neuron models, the inter-spike interval (ISI) distribution $f(\tau)$ can be written in terms of the PSTH as

\small
\begin{equation}
\label{eq20}
f(\tau|s) = \frac{1}{N} \int_{0}^{T-\tau} {\rm d}t~P[t|s]~P[t+\tau|s]~\exp \left\{-\int_{t}^{t+\tau}P[t'|s]~{\rm d}t'\right\},
\end{equation}

\normalsize
\noindent where $T$ is the total length of the recording, and $N$ is the total number of spikes, $N = \int_{0}^{T}P[t|s] \ {\rm d}t$. The factors $P[t|s]$ and $P[t + \tau | s]$ represent the probability of spike generation at times $t$ and $t + \tau$, respectively. The exponential factor is the probability that no spike is fired inside the interval $(t, t + \tau)$. It is important to keep in mind that Eq.~(\ref{eq20}) is only valid within the framework of Poisson neurons, that is, cell models in which the firing probability only depends on the incoming stimulus. In particular, all the effects of previous cellular activity (refractoriness, adaptation) are neglected. Strictly speaking, dynamical neuron models cannot be reduced to a Poisson formalism, although in some cases an approximate reduction is justified \citep{aviel2006}. Hence, Eq.~(\ref{eq20}) should be understood as a property of theoretical Poisson models, that is only expected to describe the behavior of real neurons for intervals $\tau$ that are large compared to the cellular processes that regulate after-spike
hyperpolarization and repolarization.

\indent If $f^{s}$ is the ISI distribution of a constant stimulus $s(t) \equiv s$, then

\small
\begin{equation}
\label{eq21}
f^{s}(\tau) = \frac{T-\tau}{T}~P_{0}(s)~\exp[-\tau~P_{0}(s)]
\end{equation}

\normalsize
\noindent which, in the limit $T \rightarrow \infty$ tends to the classical expression for Poisson spike trains $f^{s}(\tau)$.
Hereafter, we assume that $T \gg \tau$, so the superior integration limit in Eq.~(\ref{eq20}) is replaced by $T$.

\indent Decomposing the PSTH using Eq.~(\ref{eq7}) and keeping only the lowest order term, the ISI distribution reads

\small
\begin{equation}
\label{eq22}
f(\tau|s) \approx \frac{1}{N} \int_{0}^{T} {\rm d}t~P_{0}[s(t)]~P_{0}[s(t+\tau)]~\exp \left\{-\int_{t}^{t+\tau} P_{0}[s(t')]~{\rm d}t'\right\}
\end{equation}

\normalsize
\indent To further explore Eq.~(\ref{eq22}), we separately consider three different regimes, depending on how $\tau$ compares to the memory time constant of the system, $\tau_{\rm m}$.

\subsection{The ISI distribution for small $\tau$}

\indent When the ISI distribution is evaluated at a time $\tau$ that is much smaller than the typical time scale of stimulus
fluctuations, the inner integral of Eq.~(\ref{eq22}) may be replaced by $\exp\left\{- ~\tau P_{0}[s(t)]\right\}$. In addition, $P[s(t + \tau)] \approx P[s(t)]$. With these approximations,

\small
\begin{equation}
\label{eq23}
f(\tau|s) \approx \frac{1}{N}\int_{0}^{T}{\rm d}t~P_{0}[s(t)]~f^{s(t)}(\tau).
\end{equation}

\begin{figure}[ht]
\begin{center}
\includegraphics[width=0.4\textwidth]{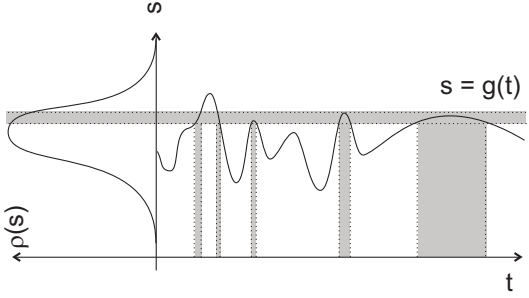} \caption{\label{f8} {\bf Construction of the distribution of stimuli $\rho(s)$ from the temporal evolution $s(t)$}. One option is to integrate the signal $g(t)$ in time. Another possibility, is to collapse the signal on the vertical axis, giving a weight to the $(s, s + {\rm d}s)$ proportional to the total time for which the signal $g(t)$ falls inside this range. If ${\rm d}s$ is small, this time is a sum of terms $|{\rm d}s/{\rm d}t|^{-1}$, one such term for each point where $g(t) = s$}
\end{center}
\end{figure}

\normalsize
\indent The ISI distribution becomes a temporal average of the constant-stimulus distribution $f^{s}(\tau)$, with a weight
function that is proportional to the PSTH. Moreover, for stationary stimuli, the temporal integral of Eq.~(\ref{eq23}) can be transformed into an integration in stimuli. In Fig.~\ref{f8} we show how to construct the stimulus density $\rho(s)$ from a time-dependent signal $s = g(t)$, by counting the fraction of stimuli that fall in the interval $(s, s + {\rm d}s)$

\small
\begin{eqnarray}
\label{eq24}
\rho(s) & = & \int_{0}^{T} P(s, t) ~ {\rm d}t = \frac{1}{T} \int_{0}^{T} \delta\left[s - g(t)\right] ~ {\rm d}t \nonumber\\         & = & \frac{1}{T} \int_{0}^{T} \delta\left[t - g^{-1}(s)\right] ~ \left|\frac{{\rm d}g}{{\rm d}t} \right|^{-1} ~ {\rm d}t \nonumber \\
        & = & \frac{1}{T} \sum_{t_{i} / g(t_{i}) = s} \left| \frac{{\rm d}g}{{\rm d}t} \right|^{-1}_{t_{i}}.
\end{eqnarray}

\normalsize
\indent The derivative ${\rm d}g/{\rm d}t$ transforms the temporal differentials into stimulus differentials, i.e., accounts for the variable widths of the shaded temporal intervals in Fig.~\ref{f8}. The points $t_{i}$ are all the times for which $g(t) = s$, and are located at the center of the vertical shaded areas of Fig.~\ref{f8}. The times for which the stimulus varies slowly, hence, have a large contribution to $\rho(s)$. Therefore, in the stimulus domain, Eq.~(\ref{eq23}) becomes

\small
\begin{eqnarray}
\label{eq25}
f(\tau | s) & = & \frac{1}{N}  \int {\rm d}s \left[ \sum_{t_{i} / g(t_{i}) = s} \left|\frac{{\rm d}g}{{\rm d}t} \right|^{-1}_{t_i} \right] ~ P_{0}(s) ~ f^{s}(\tau) \nonumber \\
            & = & \frac{T}{N} \int \rho(s) ~ P_{0}(s) ~ f^{s}(\tau) ~ {\rm d}s.
\end{eqnarray}

\normalsize
\indent Here, the first equality results from changing the integration variable from $t$ to $s$, and collecting together
all the time points $t_{i}$ that correspond to one particular $s$. The second equality derives from the definition of $\rho(s)$ in Eq.~(\ref{eq24}). Equation~(\ref{eq25}) is useful to calculate the behavior of $f(\tau|s)$ for small $\tau$, without requiring any knowledge of the temporal evolution of the PSTH. Only the activation curve $P_{0}(s)$ and the stimulus distribution $\rho(s)$ are needed. In particular, two different stimuli with the same distribution $\rho(s)$ but very different temporal characteristics (an example is shown below) lead to the same $f(\tau | s)$.

\indent In order to obtain systematic corrections to Eq.~(\ref{eq23}) when $\tau$ becomes comparable to the time scale of
stimulus fluctuations, the functions $P[t|s]$ in Eq.~(\ref{eq20}) should be replaced by several terms of their Taylor
expansion (not just the first).

\subsection{The ISI distribution for large $\tau$}

\indent If the ISI distribution is evaluated for an interval $\tau$ that is much longer than the typical time scale of stimulus variations, then the inner integral of Eq.~(\ref{eq20}) contains many stimulus fluctuations. The number of spikes generated in the interval $(t, t + \tau)$ can thus be assumed to be equal to $N ~ \tau / T$, and the exponential factor may be replaced by $\exp(- \tau ~ N/T)$. Hence,

\small
\begin{equation}
\label{eq26}
f(\tau|s) \approx \frac{1}{N} ~ \exp\left(- \frac{N}{T} ~ \tau \right) ~ \int_{0}^{T}{\rm d}t~P_{0}[s(t)] ~ P_{0}[s(t+\tau)].
\end{equation}

\normalsize
\indent The ISI distribution is exponential in $\tau$, and proportional to the autocorrelation of the spike train.

\subsection{The ISI distribution for intermediate $\tau$}

\indent When $\tau$ is comparable to the typical time scale of stimulus fluctuations, the ISI distribution must be calculated with the full Eq.~(\ref{eq20}) or keeping more terms for $P[t|s]$.

\subsection{Examples for the linear neuron model}

\subsubsection{Deterministic input currents}

{\bf Periodic square wave}\\

\begin{figure*}
 \centering
   \begin{minipage}[t]{.2\textwidth}
     \centering
     \par\vspace{0pt}
     \caption{\label{f9} {\bf Examples of ISI distributions obtained by a linear ON cell, driven by slow deterministic and stochastic stimuli}. Left panels: stimulus distributions $\rho(s)$. Middle panels: temporal evolution of the stimulus. Right panels: ISI distributions obtained from Eq.~(\ref{eq25}) for short $\tau$ (red continuous line) and histograms obtained by numerical simulation of the dynamical system (stairs in black line, with grey-shaded area). The dashed line serves for comparison with the exponential distribution, corresponding to a constant stimulus of the same mean value. Square, triangular and sinusoidal waves: $\Delta = 0.025 ~{\rm [stim]}$, $T = 200 ~{\rm ms}$, $h_{0} = 100 ~{\rm Hz}$, $H_{1}^{0} = 2.506 ~{\rm ms^{-1} [stim]^{-1}}$. Gaussian signals (d-e) have ${\rm SD = 0.01 ~[stim]}$}
   \end{minipage}%
   \begin{minipage}[t]{.85\textwidth}
     \centering
     \par\vspace{0pt}
     \includegraphics[width=0.85\textwidth]{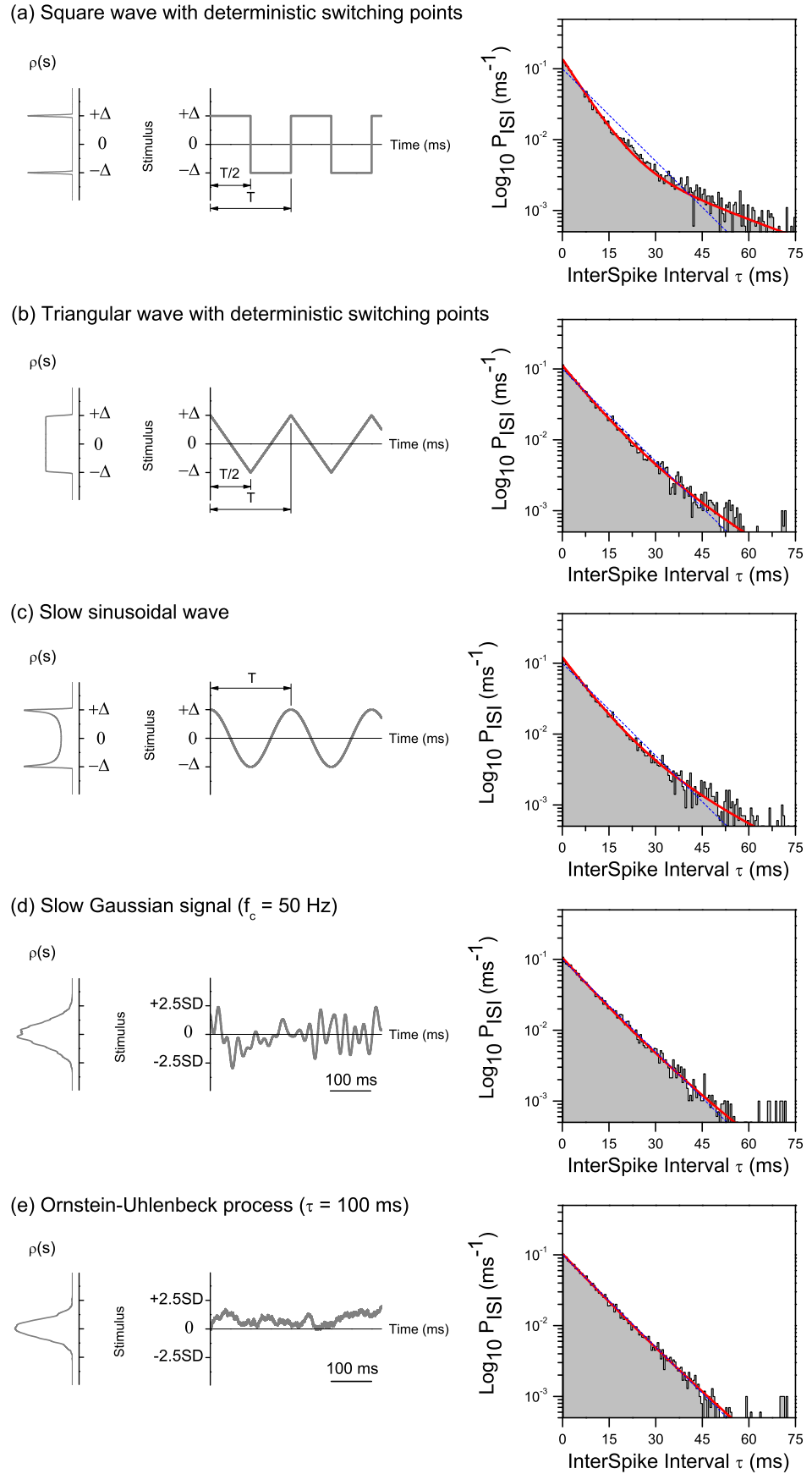}
   \end{minipage}
\end{figure*}

\indent In Fig.~\ref{f9} A, we show a square stimulus of amplitude $\Delta$ and period $T$. This is a binary stimulus, whose
distribution $\rho(s) = [\delta(s - \Delta) + \delta (s + \Delta)]/2$ is depicted on the left. Integrating Eq.~(\ref{eq25}),
we obtain

\small
\begin{equation}
\label{eq27}
f(\tau) = \frac{1}{2h_0} \left[ \mathcal{H}_{+}^2 \exp\left( - \tau~\mathcal{H}_{+}\right) + \mathcal{H}_{-}^2 \exp\left( - \tau~\mathcal{H}_{-}\right) \right],
\end{equation}

\normalsize

\noindent where $\mathcal{H}_{\pm} = h_{0} \pm H_{1}^{0}~\Delta$. In Fig.~\ref{f9} A, we see that the ISI distribution corresponding to a slow square-wave stimulus has a higher peak and a longer tail than the purely exponential distribution of decay rate $h_{0}$ obtained for a constant stimulus of the same mean. This is because there are two exponentials in Eq.~(\ref{eq27}). One of them has a slower decay rate than $h_{0}$, and is responsible for the long tail of the distribution, at large $\tau$. The other exponential decays faster than $h_{0}$, and dominates for small $\tau$. Notice that although Eq.~(\ref{eq25}) was derived in the limit of short $\tau$, the red curve reproduces the numerical data up to $\tau$ values that are $6$ times larger than the mean ISI, in this case, $10$ msec.\\

{\bf \noindent Periodic triangular wave}\\

\indent We now consider a triangular stimulus of amplitude $\Delta$ and period $T$, as in Fig.~\ref{f9} B. In this case, $\rho(s) = 1/2\Delta$, for $s \in [-\Delta, \Delta]$. By integrating Eq.~(\ref{eq25}), we get

\small
\begin{eqnarray}
\label{eq28}
f(\tau)  =  \frac{1}{2 \Delta H_{1}^{0} h_{0} \tau} \big[ && \left( \mathcal{H}_{-}^{2} + 2 \mathcal{H}_{-} / \tau + 2/\tau^{2} \right) \exp \left( -\tau ~ \mathcal{H_{-}}\right)  - \nonumber \\
 - && \left( \mathcal{H}_{+}^{2} + 2 \mathcal{H}_{+} / \tau + 2/\tau^{2} \right) \exp \left( -\tau ~ \mathcal{H_{+}}\right) ~ \big].
\end{eqnarray}

\normalsize
\indent This ISI distribution also contains two exponentials with slightly different decay times, producing a higher peak at the origin, and a longer tail at large $\tau$ values, as shown in Fig.~\ref{f9} B. The effect is less noticeable as with the square wave, given that this stimulus spends less time in the extreme values.\\

{\bf \noindent Sinusoidal wave}\\

\indent For a sinusoidal stimulus of Fig.~{\ref{f9}} C, the stimulus density is $\rho(s) = (\pi \Delta \sqrt{1 - s^{2} /
\Delta^{2}})^{-1}$, for $s \in [-\Delta, \Delta]$. The numerical integration of Eq.~(\ref{eq25}) is depicted in the right panel of Fig.~\ref{f9} C. Also in this case the obtained distribution differs from the exponential distribution, by displaying a higher peak at zero, and a longer asymptotic tail. Once again the approximation for small $\tau$ values correctly reproduces the entire range of ISIs.

\subsubsection{Stochastic input currents}

{\bf The Gaussian stimulus}\\

\indent For a Gaussian stimulus of standard deviation $\sigma$ and cutoff frequency $f_{\rm c}$

\small
\begin{equation}
\label{eq29}
f(\tau) = \frac{1}{\tau_{0}} ~ \{ \left[ \sigma H_{1}^{0} (\tau-\tau_{0}) \right]^{2} + 1 \} ~ \cdot ~\exp\left[-\frac{h_{0}^{2}}{4}(\tau-\tau_{0})^{2}\right]
\end{equation}

\normalsize
\noindent where $\tau_{0} = h_{0} / (\sigma H_{1}^{0})^{2}$. The ISI distribution for small $\tau$ does not depend on the cutoff frequency of the stimulus, that is, on its temporal correlations. These correlations only participate in determining the range of validity of Eq.~(\ref{eq29}). In the right panel of Fig.~\ref{f9} D, however, we see that the range of validity
of Eq.~(\ref{eq29}) extends throughout the whole range of ISIs. Hence, the ISI distribution of slow Gaussian signals is independent of the cutoff frequency.

\indent In Eq.~(\ref{eq29}), the ISI distribution has a quadratic dependence on $\tau$ (both in the factor and in the exponential term). The conditions of validity of the approximation ($h_{0} \gg |\sigma H_{1}^{0}|$) imply that $\tau_{0}$ must be large. In this limit, Eq.~(\ref{eq29}) reduces to a purely exponential function, visually indistinguishable from the one obtained for constant stimulation, as shown in the right panel of Fig.~\ref{f9}~D.

\indent In the limit of large $\tau$, the ISI distribution reads

\small
\begin{equation}
\label{eq30}
f(\tau) \to \frac{\exp\left(- \tau ~ N / T\right)}{N} \left[ h_{0}^{2} T + \left( H_{1}^{0} \right)^{2} \int_{0}^{T} s(t)~s(t + \tau) ~ {\rm d}t \right].
\end{equation}

\normalsize
\indent The integral in Eq.~(\ref{eq30}) is the autocorrelation of the stimulus. For white Gaussian stimuli with
cutoff frequency $f_{\rm c}$, the integral can be calculated analytically, and

\small
\begin{equation}
\label{eq31}
f(\tau) \to \frac{\exp\left(-h_{0}~\tau \right)}{h_{0}} ~ \left[ h_{0}^{2} + \left( H_{1}^{0} \right)^{2} \frac{\sigma^{2}}{2 \pi } ~ \frac{\sin(2 \pi f_{\rm c} ~\tau)}{f_{\rm c}~\tau} \right],
\end{equation}

\normalsize
\noindent where we have replaced $N/T$ by $h_{0}$.

\indent The tail of the ISI distribution oscillates with the cutoff frequency of the stimulus. These oscillations, however, are masked by the rapid decay imposed by the exponential term, and the $\tau$ in the denominator of Eq.~(\ref{eq31}). They are therefore invisible in Fig.~\ref{f9} D.\\

{\bf \noindent The Ornstein-Uhlenbeck process}\\

\indent Consider an Ornstein-Uhlenbeck process with a correlation time $\lambda$ defined by the equation

\small
\begin{equation}
\label{eq32}
\lambda ~ {\rm d}s/{\rm d}t = - s(t) + \sigma ~ \sqrt{\lambda} ~ \xi(t),
\end{equation}

\normalsize
\noindent where $\xi(t)$ represents a white noise realization of zero mean and unit variance. The distribution of the stimulus is a Gaussian function of zero mean and standard deviation $\sigma$. Therefore, the ISI distribution for small $\tau$ coincides with Eq.~(\ref{eq29}).

\indent In order to obtain the ISI distribution for large $\tau$, we write Eq.~(\ref{eq26}) as

\small
\begin{eqnarray}
\label{eq33}
f(\tau) & \to & \frac{1}{h_{0}} ~ \exp\left(- \frac{N}{T}~\tau \right) ~ \left[ h_{0}^{2}  + \left(H_{1}^{0}\right)^{2} \frac{1}{T} \int_{0}^{T}{\rm d}t ~ s(t) ~ s(t + \tau) \right] \nonumber \\
& = & \frac{\exp\left(- h_{0}~\tau \right)}{h_{0}} \left[ h_{0}^{2} + (H_{1}^{0})^{2} ~ \sigma^{2} ~ {\rm e}^{-\tau/\lambda} / 2\right].
\end{eqnarray}

\normalsize
\indent As in the previous case, this distribution is dominated by the exponential term (see Fig.~\ref{f9} E).

\section{Conclusions}

\indent We derived analytical expressions for the PSTH and the ISI distribution of arbitrary Poisson neurons, when driven with slow stimuli. Our results depend on a combination of intrinsic neuronal properties (captured by the $H_k^{j_1, \dots j_k}$), and the external stimulus (represented by $\lambda_j$). In the particular case of linear neuron models, the PSTHs of ON and OFF cells are proportional to the delayed stimulus. The constant of proportionality depends on the total area under the linear filter, whereas the delay is solely determined by its center of mass. For biphasic linear neurons, the PSTH is proportional to the delayed stimulus derivative. The constant of proportionality and the delay can be easily determined by performing an elementary integration of functions involving the linear filter.

\indent When driven with small-amplitude stimuli, LNL neuron models can be linearized around their spontaneous rate. The resulting PSTH is formally equivalent to the one that would be obtained from an effective linear model. For larger stimulus amplitudes, however, upward and downward stimulus deflections are not symmetrically represented in the PSTH. This asymmetry is the footprint of the influence of the curvature of the activation curve. Depending on whether the activation curve operates below or above its inflection point, positive or negative stimulus curvatures are selectively favored.

\indent These results might be useful in several applications. They could be useful, for example, in the analysis of the information obtained by a population of LNL neurons. Several studies have aimed to determine the optimization principle that underlies the way in which retinal receptive fields tile the visual space \citep{devries1997,vincent2005,segev2006,borghuis2008,gauthier2009}. These studies focus on the population distribution of the spatial properties of receptive fields. In this context, it would be interesting to understand the population distribution of also the temporal properties of receptive fields. These properties involve specific time scales (slow, medium and fast), where different time-scales are represented with different frequencies in ON, OFF, biphasic ON and biphasic OFF cells \citep{segev2006,gollisch2008a}. There is presumably also an optimization principle that has guided
the evolution towards the particular distribution of temporal receptive field properties found in each species, probably dependent on the statistics of natural images. Our study provides a simple picture of how the PSTH of a LNL model depends on the incoming signals. Therefore, it should be useful to determine the type and amount of information that is encoded by a population of neurons characterized by a specific distribution of temporal receptive field properties.

\indent In the second part of the paper, we derive the dependence of the ISI distribution on both neuronal and stimulus properties. When a Poisson cell is driven with a constant stimulus, an exponential distribution is found. Here we showed that for variable stimuli, the ISI distribution can be obtained by weighting the exponential distribution with the stimulus density. Strictly speaking, this result is only valid in the limit of small ISI. However, numerical evaluation of the ISI distribution for the case of slow stimuli showed that this expression reproduces the whole range of ISIs. Under this approximation, the ISI distribution is solely determined by the stimulus distribution $\rho(s)$, independently of the
stimulus' temporal properties. In particular, the ISI distribution obtained from a periodic stimulus does not depend on the stimulus frequency. In the same way, all Gaussian stimuli give rise to the same ISI distribution, no matter the cutoff frequency. Indeed, Gaussian stimuli and Ornstein-Uhlenbeck processes produce also the same distribution. The obtained densities show deviations from the exponential behavior. Specifically, the probability density of both small and large ISIs is incremented, at the expense of the probability density of intermediate ISIs. This result was more visible for deterministic stimuli than for stochastic signals. Within the class of deterministic stimuli, the largest effect was observed in the case of the square wave. Presumably, those input currents whose distribution $\rho(s)$ differs most markedly from the constant case (where $\rho(s) = \delta(s - s_0)$, for a given $s_0$) are the ones whose ISI distribution is most strikingly different from the exponential function.

\indent In \citep{urdapilleta2009}, we derived the ISI distribution of a perfect integrate-and-fire model neuron, when
driven with slow stimuli. The analytical expression coincided with Eq.~(\ref{eq23}), implying that the result that we obtained here for the particular case of a Poisson neuron is also applicable to other dynamical neuron models.

\indent We have focused on slow driving signals. As mentioned in the introduction, there is a myriad of modulating processes in the brain that operate in a slow time scale. In sensory systems, the biological relevance of extremely fast stimuli is questionable, since it has been argued \citep{atick1992} that the temporal spam of receptive fields has evolved as to average out fast stimulus fluctuations, often contaminated with high levels of noise. However, a description of the effect of stimuli whose typical time scales are comparable to those of the receptive fields would be highly desirable. This would mean to extend the present results to the case where the stimulus fluctuations are comparable to $\tau_{\rm m}$. Presumably these time scales are the most relevant for accurate behavioral performance.

\onecolumn
\section{Appendix}

\indent Starting from Eq.~(\ref{eq1}) and replacing $(-\infty,+\infty)$ integration limits by $(t-\tau_{\rm m},t)$ we  obtain

\begin{eqnarray}
\label{eq2}
 P[t|s] = h_{0} & + & \int_{t - \tau_{\rm m}}^{t} {\rm d}t_{1}~h_{1}(t-t_{1})~s(t_{1}) +  \int_{t - \tau_{\rm m}}^{t} {\rm d}t_{1} \int_{t - \tau_{\rm m}}^{t} {\rm d}t_{2}~h_{2}(t-t_{1},t-t_{2})~s(t_{1})~ s(t_{2}) +  \dots + \nonumber \\
 & + & \int_{t-\tau_{\rm m}}^{t} {\rm d}t_{1} \dots \int_{t-\tau_{\rm m}}^{t} {\rm d}t_{k}~h_{k}(t-t_{1},
\dots,t-t_{k})~s(t_{1})\dots s(t_{k}) + \dots
\end{eqnarray}

\indent Replacing the Taylor expansion of the stimulus, Eq.~(\ref{eq3}), in Eq.~(\ref{eq2}),

\begin{eqnarray}
\label{eq4}
P[t|s] & = & h_{0} + \sum_{k = 1}^{+\infty} \int_{t-\tau_{\rm m}}^{t} {\rm d}t_{1} \dots \int_{t-\tau_{\rm m}}^{t} {\rm d}t_{k} ~ h_{k}(t-t_{1},\dots,t-t_{k}) \prod_{\ell = 1}^{k}~s(t_{\ell}) \nonumber \\
       & = & h_{0} + \sum_{k = 1}^{+\infty} \int_{t-\tau_{\rm m}}^{t} {\rm d}t_{1} \dots \int_{t-\tau_{\rm m}}^{t} {\rm d}t_{k} ~  h_{k}(t-t_{1},\dots,t-t_{k}) \prod_{\ell = 1}^{k} \sum_{j_{\ell} = 0}^{+\infty} \frac{\lambda_{j_{\ell}}}{j_{\ell}!}(t_{\ell}-t)^{j_{\ell}} \nonumber \\
       & = & h_{0} + \sum_{k = 1}^{+\infty} \int_{t-\tau_{\rm m}}^{t} {\rm d}t_{1} \dots \int_{t-\tau_{\rm m}}^{t} {\rm d}t_{k} ~ h_{k}(t-t_1,\dots,t-t_{k}) \sum_{\{j_{1},\dots,j_{k}\}} \prod_{\ell = 1}^{k}\frac{\lambda_{j_{\ell}}}{j_{\ell}!}(t_{\ell}-t)^{j_{\ell}},
\end{eqnarray}

\noindent where the last sum runs over all the sets of $k$ non-negative numbers $j_{1},\dots, j_{k}$. We now rearrange the sum in $\{j_{1},\dots,j_{k}\}$ in increasing order of the product of derivatives. To that end, we group together the terms with the same value of the sum $r = \sum_{i = 1}^{k} j_{i}$, that is,

\begin{eqnarray}
\label{eq5}
P[t|s] & = & h_{0} + \sum_{k = 1}^{+\infty} \int_{t-\tau_{\rm m}}^{t} {\rm d}t_{1} \dots \int_{t-\tau_{\rm m}}^{t} {\rm d}t_{k} ~ h_{k}(t-t_{1},\dots,t-t_{k}) \sum_{r = 0}^{+ \infty} \sum'_{\{j_{1},\dots,j_{k}\}} \prod_{\ell = 1}^{k} \frac{\lambda_{j_{\ell}}}{j_{\ell}!}(t_{\ell}-t)^{j_{\ell}} \nonumber \\
       & = & h_{0} + \sum_{k = 1}^{+\infty} \sum_{r = 0}^{+\infty} \sum'_{\{j_{1},\dots,j_{k}\}} \prod_{\ell = 1}^{k}
\frac{\lambda_{j_{\ell}}}{j_{\ell}!} \int_{t-\tau_{\rm m}}^{t} {\rm d}t_{1} \dots \int_{t-\tau_{\rm m}}^{t} {\rm d}t_{k} ~ h_{k}(t-t_{1},\dots,t-t_{k}) \prod_{i =1}^{k} (t_{i}-t)^{j_{i}} \nonumber\\
       & = & h_{0} + \sum_{r = 0}^{+\infty} \sum_{k = 1}^{+\infty} \sum'_{\{j_{1},\dots,j_{k}\}} \prod_{\ell = 1}^{k}
\frac{\lambda_{j_{\ell}}}{j_{\ell}!} ~ H_{k}^{j_{1},\dots,j_{k}},
\end{eqnarray}

\noindent where the primed sum $\sum'_{\{j_{1},\dots,j_{k}\}}$ runs over non-negative integers $\{j_{1},\dots,j_{k}\}$ that sum up to $r$, and $H_{k}^{j_{1},\dots,j_{k}}$ was defined in Eq.~(\ref{eq6}). The different terms in Eq.~(\ref{eq7}) correspond to the different $r$-values in Eq.~(\ref{eq5}).

\twocolumn


\begin{thebibliography}{}

\bibitem[Albrecht and Hamilton(1982)]{albrecht1982}
Albrecht DG, Hamilton DB (1982) Striate cortex of monkey and cat: Contrast response function. J Neurophysiol 48(1): 217-237\\

\bibitem[Albright(1984)]{albright1984}
Albright TD (1984) Direction and orientation selectivity of neurons in visual area MT of the macaque. J Neurophysiol 52(6): 1106-1130\\

\bibitem[Atick(1992)]{atick1992}
Atick JJ (1992) Could information theory provide an ecological theory of sensory processing?. Network 3:213-251\\

\bibitem[Aviel and Gerstner(2006)]{aviel2006}
Aviel Y, Gerstner W (2006) From spiking neurons to rate models: A cascade model as an approximation to spiking neuron models with refractoriness. Phys Rev E 73: 051908/1-10\\

\bibitem[{Benda and Herz(2003)}]{benda2003}
Benda J, Herz AVM (2003) A universal model for spike-frequency adaptation. Neural Comput 15(11): 2523-2564\\

\bibitem[Borghuis et al.(2008)]{borghuis2008}
Borghuis BG, Ratliff CP, Smith RG, Sterling P, Balasubramanian V (2008) Design of a Neuronal Array. J Neurosci 28(12): 3178-3189\\

\bibitem[Brenner et al.(2000)]{brenner2000}
Brenner N, Bialek W, de Ruyter van Steveninck R (2000) Adaptive rescaling maximizes information transmission. Neuron 26: 695-702\\

\bibitem[De Vries and Baylor(1997)]{devries1997}
De Vries SH, Baylor DA (1997) Mosaic Arrangement of Ganglion Cell Receptive Fields in Rabbit Retina. J Neurophysiol 78: 2048-2060\\

\bibitem[Doya(2002)]{doya2002}
Doya K (2002) Metalearning and neuromodulation. Neural Networks 15: 495-506\\

\bibitem[Ermentrout(1998)]{ermentrout1998}
Ermentrout B (1998) Linearization of f-I curves by adaptation. Neural Comput 10(7): 1721-1729\\

\bibitem[Fontanini and Bower(2006)]{fontanini2006}
Fontanini A, Bower JM (2006) Slow-waves in the olfactory system: an olfactory perspective on cortical rhythms. Trends Neurosci 29(8): 429-437\\

\bibitem[Gauthier et al.(2009)]{gauthier2009}
Gauthier JL, Field GD, Sher A, Greschner M, Shlens J, Litke AM, Chichilnisky EJ (2009) Receptive fields in primate retina
are coordinated to sample visual space more uniformly. PLoS Biol 7(4): e1000063\\

\bibitem[Georgopoulos et al.(1982)]{georgopoulos1982}
Georgopoulos AP, Kalaska JF, Caminiti R, Massey JT (1982) On the relations between the direction of two-dimensional arm movements and cell discharge in primate motor cortex. J Neurosci 2(11): 1527-1537\\

\bibitem[Gollisch and Meister(2008a)]{gollisch2008a}
Gollisch T, Meister M (2008) Rapid neural coding in the retina with relative spike latencies. Science 319: 1108-1111\\

\bibitem[Gollisch and Meister(2008b)]{gollisch2008b}
Gollisch T, Meister M (2008) Modeling convergent ON and OFF pathways in the early visual system. Biol Cybern 99: 263-278\\

\bibitem[Lesica and Grothe(2008)]{lesica2008}
Lesica NA, Grothe B (2008) Efficient temporal processing of naturalistic sounds. PLoS ONE 3(2): e1655\\

\bibitem[Nagel and Doupe(2006)]{nagel2006}
Nagel KI, Doupe AJ (2006) Temporal processing and adaptation in the songbird auditory forebrain. Neuron 51: 845-859\\

\bibitem[Rust et al.(2005)]{rust2005}
Rust NC, Schwartz O, Movshon JA, Simoncelli EP (2005) Spatiotemporal elements of macaque V1 receptive fields. Neuron 46: 945-956\\

\bibitem[Saper(2000)]{saper2000}
Saper CB (2000) Brain stem modulation of sensation, movement and consciousness. In: Kandel ER, Schwartz JH, Jessell TM (eds) Principles of neural science, 4th edn, McGraw-Hill, New York, pp 889-909\\

\bibitem[Schwartz et al.(2006)]{schwartz2006}
Schwartz O, Pillow JW, Rust NC, Simoncelli EP (2006) Spike-triggered neural characterization. J Vis 6: 484-507\\

\bibitem[Segev et al.(2006)]{segev2006}
Segev R, Puchalla J, Berry MJII (2006) Functional organization of ganglion cells in the salamander retina. J Neurophysiol 95: 2277-2292\\

\bibitem[Toth et al.(1997)]{toth1997}
T\'{o}th K, Freund TF, Miles R (1997) Disinhibition of rat hippocampal pyramidal cells by GABAergic afferents from the septum. J Physiol 500(2): 463-474\\

\bibitem[Urdapilleta and Samengo(2009)]{urdapilleta2009}
Urdapilleta E, Samengo I (2009) Quasistatic approximation of the interspike interval distribution of neurons driven by time-dependent inputs. Phys Rev E 80: 011915/1-7\\

\bibitem[Vincent et al.(2005)]{vincent2005}
Vincent BT, Baddeley RJ, Troscianko T, Gilchrist ID (2005) Is the early visual system optimised to be energy efficient?. Netw: Comput Neural Syst 16(2/3): 175-190\\

\bibitem[Wilson et al.(2004)]{wilson2004}
Wilson CJ, Weyrick A, Terman D, Hallworth NE, Bevan MD (2004) A model of reverse spike frequency adaptation and repetitive firing of subthalamic nucleus neurons. J Neurophysiol 91: 1963-1980\\

\end{thebibliography}
\end{document}